\documentclass[preprint,aps,tightenlines,byrevtex,showpacs,nofootinbib]{revtex4}

\usepackage{amssymb,amsmath}

\def\Mp{M_p}
\def\mp{ {\bar{M}}_{p} }
\def\Mf{M_f}

\def\pn{\phi_{\vec{n}}}
\def\mn{m_{\vec{n}}}
\def\s{\sqrt{s}}
\def\d{\displaystyle}
\def\M{\mathcal{M}}
\def\L{\mathcal{L}}
\def\O{\mathcal{O}}
\def\sig{{\sigma}_{\Gamma}}
\def\4!{\!\!\!\!}
\def\2!{\!\!}
\def\simge{\mathrel{
   \rlap{\raise 0.511ex \hbox{$>$}}{\lower 0.511ex \hbox{$\sim$}}}}
\def\simle{\mathrel{
   \rlap{\raise 0.511ex \hbox{$<$}}{\lower 0.511ex \hbox{$\sim$}}}}
\def\v{\vphantom{ \bigl[\bigr] }}

\def\n{\left| \vec{n} \right|}

\begin{document}

\title{ A Problem in Virtual Graviton exchanges 
             in Flat Large Extra Dimensions}

\author{Haruka Namatame}
\email{haruka@phys.metro-u.ac.jp}
\affiliation{Department of Physics, Tokyo Metropolitan University,
Minami-Osawa, Hachioji, Tokyo 192-0397, Japan}

\begin{abstract}
It is pointed out in a class of models with large extra dimensions 
that the cross section of processes with
virtual Kaluza-Klein graviton exchanges becomes either much smaller
or much larger by many orders of magnitude than 
what is expected from that of the on-shell production 
of the Kaluza-Klein gravitons.
We demonstrate how the problem arises using a toy model.
The cause of this new problem lies in the fact that we do not have 
momentum conservation in the extra dimensions.
To search for the signal of the large extra dimensions with
high energy collider experiments,  
we need more care in interpreting the earlier results on 
the cross sections of these processes.

\end{abstract}

\maketitle

\newpage

\section{Introduction}
\label{introduction}

In order to solve the hierarchy problem between the electroweak scale
and the Planck scale, extra space dimensions can play an important role. 
The scenario of large extra dimensions, 
which was proposed by Antoniadis 
\cite{Antoniadis:1990ew,Antoniadis:1992fh,Antoniadis:1993jp,Antoniadis:1998ig}
 is very simple.
In this scenario, we assume that
the Standard Model (SM) particles live on a three-dimensional hypersurface
and the only graviton lives in the whole higher dimensional space.

The simplest and explicit model with this scenario is the one by 
Arkani-hamed, Dimopoulos, and Dvali (ADD)
\cite{ArkaniHamed:1998rs,ArkaniHamed:1998nn}.
ADD have introduced a $\delta$-dimensional flat extra space which
is compactified on a $\delta$-dimensional torus with common radius $R$.
In this model, the Planck scale $\Mp$ in our four-dimensional space time
is given by the fundamental Plank scale $\Mf$ in the whole $4 +
\delta$-dimensional spacetime and radius $R$ of the extra dimensions as
\begin{eqnarray}
M_{p}^{2} = 8 \pi R^{\delta}M_{f}^{2+ \delta}.
\end{eqnarray}
If $R$ is large enough, we can take the fundamental Planck scale
$\Mf$ to be a few TeV, which naturally explains the hierarchy between the
energy scale of electroweak interaction and that of
gravitational interaction.

Note that in the ADD model, 
localization of the SM particles on the three-dimensional 
hypersurface is described by a delta function.
In this way, the three-dimensional hypersurface is treated as a complete
rigid-body and the momentum in the extra dimensions is not conserved
 in the interaction between the higher dimensional graviton 
and the SM particles. The hypothesis of the complete rigid-body 
hypersurface leads to a explicit breaking of the translational
symmetry in the extra dimensions \cite{Bando:1999di}.

Many massive gravitons, the Kaluza-Klein excitation modes of the 
graviton (KK gravitons), are contained in the four-dimensional 
effective theory of the ADD model. 
They can decay into the light SM particles.
 We can test the ADD model by searching for
these massive gravitons directly or indirectly.
The stringent constraint comes from the supernova SN1987a
\cite{Cullen:1999hc,Barger:1999jf}. 
Using this constraint, it was shown that $\delta$ has to be larger than two
in order for the fundamental Plank scale $\Mf$
to be a few TeV.
In the rest of this paper, we will assume that $\delta$ 
satisfies this constraint.

To test the ADD model at collider experiments, two types of the process
have been studied. One is the real KK graviton emission process,
 and the other one is the virtual KK graviton exchange process. 
In case of a hadron collider, dominant channels are 
the real KK graviton emission process
$pp \rightarrow {\rm jet} + ({\rm missing})$ 
\cite{Mirabelli:1998rt,Vacavant:2001sd,Weng:2006bc},
 and the virtual KK graviton exchange process 
$pp \rightarrow l\bar{l}, \gamma \gamma$ 
\cite{Hewett:1998sn,Kabachenko:2001,Belitelov:2006}.
These studies show that the virtual KK graviton exchange processes are 
advantageous to search for signals of the ADD model, compared to  
the real KK graviton emission processes due to difficulty of identifying
the missing energy carried by the KK gravitons.

However, as we will show below, the cross section of the virtual KK
graviton exchange processes in the ADD
model have the following problems:

{\bf (I) A problem of ultraviolet divergence of momentum integration 
in extra dimensions }\\
The origin of this problem lies in the violation of the momentum conservation
in extra dimensions.
In Ref.\cite{Bando:1999di}, it was shown that we can naturally
regularize this ultraviolet divergence
by introducing a fluctuation of the three dimensional hypersurface.

{\bf (II) Breakdown of narrow width approximation}\footnote{
In Ref.\cite{Berdine:2007uv}, 
the same subject was discussed. However the main
point in  Ref.\cite{Berdine:2007uv} is that the narrow width 
approximation would be 
broken down if the mass of one of the final state particles is close to
that of the intermediate particle, because of a technical problem
in the approximation method. In the present case, as we will see
below, the problem appears irrespective of whether such condition is 
satisfied or not, and the situation is completely different from
that in Ref.\cite{Berdine:2007uv}.   
} \\
This is a new problem that we point out in this paper. 
As we will see below, the narrow width approximation 
\begin{eqnarray}
 \sigma({\rm full\ process}) = \sigma(X\ {\rm production}) 
              \times Br(X\ {\rm decay}) + \mathcal{O}(\Gamma_X/m_X)\ 
                \label{eq:NWA}
\end{eqnarray}
is not satisfied for the cross sections of the virtual KK 
graviton exchange processes.
In eq.(\ref{eq:NWA}), $\sigma({\rm full\ process})$ and 
$\sigma(X\ {\rm production})$ are the cross sections of the full process 
and production process of decaying particles, respectively, 
$\Gamma_X$ stands for the decay width of the particle,
and $Br(X\ {\rm decay})$ is the branching ratio of the 
decay channel of $X$ which is included in the full process.
Eq.(\ref{eq:NWA}) is expected to hold 
in most processes under certain conditions to be described later. 
However, as we will see below, the cross section of such processes in past 
works (for example in Ref.\cite{Han:1998sg, Giudice:1998ck}) does not satisfy 
the approximate expression in eq.(\ref{eq:NWA}).

The origin of this problem lies in lack of 
the momentum conservation
in extra dimensions. 
As we will see later, lack of the momentum conservation
in the extra dimensions causes over counting of the extra dimensional
phase space of the higher dimensional graviton or the summation over the
KK index, and we can not obtain the on-shell contributions
from the KK gravitons in the intermediate state 
(or the first term in eq.(\ref{eq:NWA})) correctly 
even if we take the decay width of the KK gravitons into account. 
Although the cause of the problem (II) is the same as that of the problem (I), 
the problem (II) is independent from the problem (I). 
This is because the problem (II) appears 
even if we put the momentum cutoff to regularize the ultraviolet
 divergence, or even if we use the prescription in Ref.\cite{Bando:1999di}.
We will see more details of this problem later.

The purpose of this paper is to clarify the problem (II).
We will show that how the problem appears in the virtual KK graviton
exchange processes.
For simplicity, we analyze these processes using a toy model
in which tensor fields are replaced by scalar fields, because
spin of the intermediate particles is not essential for
the problem (II). To show that the problem (II) arises in the 
virtual KK graviton exchanges even if we regularize the 
pole divergence with its decay width to take care of the 
on-shell contribution of the KK gravitons, we analyze the 
virtual KK graviton exchange process putting its decay width 
into the denominator of its propagator to regularize 
the on-shell pole divergence.

The rest of this paper is organized as follows. In section 2, 
we introduce the toy model, and show how we can calculate the
cross section of virtual KK graviton exchanges in this model. 
In section 3, we explain how the problem occurs in the virtual KK
 graviton exchange processes. 
Moreover, we demonstrate that the same
problem appears in the virtual KK graviton exchange processes with
three final states.
In section 4, we draw our conclusions.

\section{Virtual Kaluza-Klein graviton exchanges and a toy model}

\label{KK exchanges}

First of all, we briefly review the KK graviton and its decay
width in the ADD model. 
For more detail, see Ref.\cite{Giudice:1998ck,Han:1998sg}.

In the four-dimensional effective theory for the ADD model, 
the higher dimensional graviton whose momentum in the extra
dimensions is $\vec{n}/R$ is treated as the massive KK graviton whose
mass is $\n/R$ where $\vec{n}$ is a $\delta$-dimensional number vector.
The effective theory is valid only if the center of mass energy $\s$ and
the mass of the KK graviton $\n/R$ are much smaller than
the fundamental Planck scale, i.e., only if $\s,\  \n/R \ll \Mf$.
These conditions give the cut off momentum to regularize 
the ultraviolet divergence of momentum in extra dimensions
which derives from 
lack of dynamics for the localization of 
the SM particles in three-dimensional space dimension.
These conditions also keep perturbation theory with respect to the
gravitational vertex valid. 

 The interactions between the spin-2 KK gravitons and the SM
particles are described by the interaction Lagrangian
\begin{eqnarray}
 \L^{int}= -\frac{1}{\mp}\d\sum_{\vec{n}} 
          {G^{(\vec{n})}}_{\mu \nu}T^{\mu \nu},
 \label{eq:lag_add_int}
\end{eqnarray}
where $\mp \equiv \Mp/\sqrt{8 \pi}  \sim 10^{18}$GeV 
is the reduced planck mass, 
${G^{(\vec{n})}}_{\mu \nu}$ is a field of the KK graviton whose
mass is $\n/R$, and 
$T^{\mu \nu}$ is the energy-momentum tensor
of the SM particles. The summation of eq.(\ref{eq:lag_add_int}) extends
over all the KK modes whose mass is less than $\Mf$.
From the higher dimensional point of view, 
this summation corresponds to the momentum integration  
in the extra dimensions.

Main features of the KK gravitons in the ADD model are the following:

(i) Lack of the momentum conservation in the extra dimensions\\
In the interaction Lagrangian \ref{eq:lag_add_int}, 
the SM particles, which originally have zero momentum in the extra dimensions,
couple to the KK gravitons,
which originally have non-zero momentum in the extra dimensions, 
and the momenta in the extra dimensions are not conserved.
 
(ii) High degeneracy of the KK gravitons\\
The mass spectrum of the KK gravitons is highly degenerate.
The number of the KK gravitons which have the same mass 
is equal to a number of 
KK index $\vec{n}$ which have same length.
For example, the number of KK gravitons whose mass is 100 GeV is
$10^{14}$.

(iii) Smallness of the decay width\\
The KK gravitons can decay into standard model particles.
Main decay channels are two photons, two gluons (two light mesons) and 
two light fermions.
The total decay width of the KK gravitons $\Gamma(m)$
can be written as a following form
\begin{eqnarray}
 \Gamma(m) = N(m)\frac{m^3}{\mp^2},
\end{eqnarray}
 where $N(m)$ is a step function of the mass of KK graviton $m$ 
($ =\n/R$). The function $N(m)$ is determined by the number and phase space of 
the decay channels, and is at most one.
The total decay width of the KK gravitons are very small
compared to their masses ($\Gamma(m)/m \sim m^2/\mp^2 \ll 1$)\cite{Han:1998sg}.

For simplicity, we will discuss these features
with a toy model in which the KK gravitons and the standard model gauge
bosons are replaced by the KK scalars and massless scalars, respectively,
because the spins of the particles and the structure of the vertices are not 
important for our purposes.
 The Lagrangian of the toy model is given by
\begin{eqnarray}
 \L \ =\  \bar{\psi} i \gamma^\mu \partial_\mu \psi \ + \ 
      \frac{1}{2}\partial_\mu \chi \partial^\mu \chi \ + \ 
        \frac{1}{2} \partial_\mu \chi' \partial ^\mu \chi'\ +
      \d\sum_{\vec{n}}\left
        [ \frac{1}{2} \partial_\mu \pn \partial^\mu \pn
            - \frac{1}{2} {\mn}^2 {\pn}^2 \right] \  \nonumber \\
      +\  \d\sum_{\vec{n}}\left[ 
              -\frac{\lambda}{\v\mp} \bar{\psi} \psi \pn \chi' \right]
     \ +\ \d\sum_{\vec{n}}\left[
             - \frac{g}{\v2\mp} \pn \left( 
             \partial_\mu \chi \partial^\mu \chi+
                  \partial_\mu \chi' \partial^\mu \chi'\right) \right],
                                               \label{eq:Toy_Lag}
\end{eqnarray}
where
the fields $\pn$ are KK scalars which correspond to the KK 
gravitons whose mass is
$\n/R$, $\chi$ and $\chi'$ correspond to the standard model
 gauge bosons, $\psi$ is a massless fermion, and
the two coefficients of the interaction terms $\lambda$ and
 $g$ are real coupling constants.

The main decay modes of the KK scalars $\pn$ are
$\pn \rightarrow \chi \chi, \chi' \chi'$.
A partial decay width $\Gamma(m)$ of these modes is given by
\begin{eqnarray}
 \Gamma(m) = \frac{m^3}{128 \pi}\Biggl(\frac{g^2}{\mp^2}\Biggr)\ .
\end{eqnarray}
Another decay mode is $\pn \rightarrow \psi \bar{\psi} \chi'$, but
this decay mode is kinematically suppressed:
$\Gamma(\pn \rightarrow \psi \bar{\psi} \chi)
/\Gamma(\pn \rightarrow \chi\chi) < 10^{-3}$.

In the toy model, the KK graviton production process
$pp \rightarrow {\rm jet} + G$ is replaced by a $\pn$ production process
$\psi \bar{\psi} \rightarrow \pn\  \chi'$, and 
the virtual KK graviton exchange process
$pp \rightarrow l \bar{l}$ is by a virtual $\pn$ exchange
process $\chi \chi \rightarrow \chi' \chi'$.

\section{Breakdown of narrow width approximation}

\label{oc problem}

In this section, to see the problem of the virtual KK graviton exchange
processes, we calculate the cross section of the virtual KK scalar
exchange processes in the toy model.
To get a finite value of the cross section, we should analyze these
process putting the decay width of the KK scalars into denominator
of its propagator.
 
In the ADD model or the toy model, we can approximately treat the spectrum of 
the KK particle as a continuous one. 
Therefore for any value of $\s > 0$, 
there are KK states which can be on-shell.
This result indicates that resonant effects are important for the virtual 
KK particle exchange processes.

\subsection{Two body scattering via KK graviton}

\label{Two body}

First, let us analyze the two-body scattering process via KK scalars
$\chi \chi \rightarrow \chi' \chi'$. 
The amplitude of this process is given by
\begin{eqnarray}
 \M_{\Gamma}(\chi \chi \rightarrow \chi' \chi')= \Biggl(\frac{g}{\mp}\Biggr)^2
                     \Biggl(\frac{s}{2}\Biggr)^2
                   \d\sum_{\vec{n}}\frac{i}{s - m^2 + im\Gamma(m)}\ ,
                         \label{eq:amp2body}
\end{eqnarray}
and then the cross section of this process is given by
\begin{eqnarray}
 \sig = \frac{1}{2^9  \pi} \! \left(\frac{g }{\mp}\right)^4 \! s^3
         \left|\d\sum_{\vec{n}}\frac{i}{s - m^2 + im\Gamma(m)}
         \right|^2\ . \label{eq.cross2body}
\end{eqnarray}
To get the cross section, we have to perform the summation of
propagators,
\begin{eqnarray}
 D(\s) \equiv \d\sum_{\vec{n}}\frac{i}{s - m^2 + im\Gamma(m)}\ .
            \label{eq:sum_prop}
\end{eqnarray}
Mass splittings between different excitation modes are at most $1/R$.
If we set $\Mf$ = $1$ TeV and $\delta$ = $4$, then we have $1/R = 30$ keV, which
is very small compared to the center mass energy $\s$.
Therefore, we can approximate the summation as an intergration over mass
and we can carry it out explicitly:
\begin{eqnarray}
  D(\s) &\thickapprox&  
     \int_{0}^{\alpha \Mf}\2! dm \ \rho(m)\ \frac{i}{s - m^2 +
     im\Gamma(m)} \nonumber \\
    &=& -\frac{i}{2} S_{3}\frac{\mp^2}{\Mf^4}
         \left[ \alpha^2 + y^2 
         {\rm ln}\left[\frac{\left|y^2 - \alpha^2
			     \right|}{y^2}\right]\right] 
          + \mathcal{O}(x^2) \ ,     \label{eq.sum3}
\end{eqnarray}
where $S_3$ is the surface of three-dimensional unit sphere, 
$x = \Mf/\mp$, $y = \s/\Mf$, and 
\begin{eqnarray}
 \rho(m)dm = S_3 \frac{\mp^2}{\Mf^6}m^3 dm
\end{eqnarray} 
is the number of KK modes with its mass between $m$ and $m + dm$.
Here we put the cutoff scale to be $\alpha \Mf$, where $\alpha$
is a cutoff parameter ($\s, \n/R \leq \alpha \Mf$).
After substituting eq.(\ref{eq.sum3}) in eq.(\ref{eq.cross2body}),
we get the cross section 
\begin{eqnarray}
\sig \simeq \frac{S_3^2}{2^{11} \pi}\frac{g^4}{\Mf^8} s^3
   \left(  \alpha^2 + {\rm ln}\left[\left| y^2 - \alpha^2
			      \right|{y^2}\right] \right)^2   \ .
\end{eqnarray}
More generally, the leading term of the cross section in the case of 
the $\delta$-dimensional extra space is 
\begin{eqnarray}
 \sig \simeq \frac{1}{2^9 \pi}
      \left(\frac{S_{\delta-1}}{\delta -2}\right)^2
        \frac{g^4}{\Mf^8} s^3\alpha^{2(\delta-2)}\  
     ,\ (y^2 \leq \alpha^2) \
           \label{eq.cross2bodygen}
\end{eqnarray}
if $\delta > 2$. 
The cross section in eq.(\ref{eq.cross2bodygen}) reproduces the result 
that we get using the $i \epsilon$ method or the principal 
value integration of the propagators to regularize the 
on-shell pole divergence of KK scalars ( for more details,
see Ref. \cite{Han:1998sg})\footnote{
The leading terms in the two methods correspond the result 
assuming $\s$ and $\Gamma(m)$ in the denominator 
in eq.(\ref{eq:sum_prop}) to be zero.}.
This cross section has two problems.
Firstly, it strongly depends on the cutoff parameter
$\alpha$. Secondly, this cross section does not contain the on-shell 
contribution from KK scalars in intermediate state.
The cross section of the on-shell contribution to the 
two-body scattering via the KK scalars is given in eq.(\ref{eq.on-shell}). 
In fact, as we will see below, the on-shell contribution 
of KK scalars is canceled out upon integration over mass, and
the heaviest mode of the KK graviton, whose mass is $\alpha \Mf$, gives
the dominant contribution.
In the next subsection, to grasp the details of these problems, 
we will discuss the narrow-width approximation for the two-body 
 scattering process via the KK scalars.

\subsection{Narrow width approximation}

\label{nw app}

To discuss whether the two features of the cross section
 are physical or unphysical, let us next see whether
the narrow width approximation is satisfied in the virtual KK 
scalar exchange process.
The narrow width approximation eq.(\ref{eq:NWA}) is expected to 
be satisfied for cross sections
 of production processes of decaying particles,
in which a particle $X$ is in intermediate state,
if (i) $X$ can be on-shell in intermediate state, and 
(ii) the decay width $\Gamma_X$ of the particle is small enough
compared to its mass  $m_X$ ($\Gamma_X/m_X \ll 1$).
For example, a cross section of the muon production and its decay process 
can be approximated by a product of the real muon production
cross section and its branching ratio.

In case of the two-body scattering process in the ADD model,
 KK gravitons satisfy these two conditions:
(i) KK scalars can be on-shell for any value of $\s > 1/R$,
and (ii) the decay width of KK scalars is small enough 
($\Gamma(m)/m \sim m^2/\mp^2 \ll 1$).
Therefore, the narrow width approximation should be valid for the two-body
 scattering process via KK scalars\footnote{ 
Note that in case of the SM, we can not divide cross sections of
two-body scattering process into the two parts: a real particle production
from two bodys and its decay. 
The production cross sections are zero at almost all the values of
$\s$, because the phase space of a particle production
from two bodys in the center of mass frame is just a point.
On the other hand, in case of the ADD model,
the phase space of one KK graviton production from two SM
particles is also a point, but 
the inclusive production cross section is non-zero at any value of $\s > 0$. 
From the higher dimensional point of view, the phase space of the higher 
dimensional graviton production from two SM particles is 
not a point. Lack of the momentum conservation law in extra
 dimensions makes the volume of phase space finite.}.

In the toy model, the inclusive cross section of KK scalars production 
is given by
\begin{eqnarray}
 \sigma(\chi \chi \rightarrow \phi) =  g^2 \frac{\pi S_{\delta -1}}{2^3}
                    \frac{\s^{\delta}}{\Mf^{2+\delta}}\ .
                       \label{eq.on-shell}
\end{eqnarray} 
If $\delta \leq 6$, the cross section 
$\sig(\chi \chi \rightarrow \chi' \chi')$ of two-body scattering process
in eq.(\ref{eq.cross2bodygen}) is much smaller than 
the cross section $\sigma(\chi \chi \rightarrow \phi)$ 
in eq.(\ref{eq.on-shell})
,
because the ratio of the two cross sections is extremely smaller than one:
\begin{eqnarray}
 \frac{\sig(\chi \chi \rightarrow \chi' \chi')}
       {\sigma(\chi \chi \rightarrow \phi)}
  \sim g^2 \alpha^{2(\delta-2)}\left(\frac{\s}{\Mf}\right)^{6-\delta}\
          \ll 1 \ ,\ ( \delta \leq 6) ,
\end{eqnarray}
or 
\begin{eqnarray}
 \sig(\chi \chi \rightarrow \chi' \chi') \ll 
           \sigma(\chi \chi \rightarrow \phi) \times 
           Br(\phi \rightarrow \chi' \chi')\ ,
\end{eqnarray}
 where the branching ratio $Br(\phi \rightarrow \chi' \chi') = 1/2$.
Even though the two-body scattering process includes the real KK scalar
production process, the cross section of two-body scattering is 
much smaller than that of the real KK scalar production process,
and the narrow width approximation is not valid.

\subsection{Over counting problem}

In the previous subsection, we saw that the narrow width approximation 
is not satisfied for the virtual KK graviton exchanges.
The reason why the narrow width approximation 
fails is because of cancellation of the on-shell contribution from KK
gravitons in the intermediate state. 
The integrand
\begin{eqnarray}
 \rho(m)\frac{1}{s - m^2 + im\Gamma(m)}
\end{eqnarray}
 of the integration over mass in eq.(\ref{eq.sum3})
has a sharp peak at $m = \s$, and the value of this function is switch
from the plus peak to the minus peak at $m = \s$.
Therefore, the on-shell contribution from the peak at $m = \s$ 
is canceled out upon integration over mass, and we get
\begin{eqnarray}
  D(\s) &\simeq& \int_{0}^{\alpha \Mf}\2! dm \ \rho(m)\frac{1}{-m^2}
                 \nonumber \\
        &=& -\frac{\mp^2}{\Mf^{2+\delta}} 
               \frac{(\alpha\Mf)^{\delta -2}}{\delta-2} \nonumber \\
    &\propto&  \rho(m=\alpha\Mf) \frac{1}{-(\alpha \Mf)^2}.
\end{eqnarray} 
The cross section of the two-body scattering process is then
\begin{eqnarray}
 \sig(\chi \chi \rightarrow \chi' \chi') \propto 
     \left|\rho(m=\alpha\Mf)  \frac{1}{-(\alpha \Mf)^2} \right|^2\ .
\end{eqnarray}
This result implies that
the heaviest mode of the KK graviton,
whose mass is $\alpha \Mf$, gives
the dominant contribution to the two-body scattering,
no matter whatever value $\s (> 1/R)$ may have.
Thus the narrow width approximation is not valid, and
the cross section depends strongly on 
the cutoff parameter $\alpha$.

 We can regard this problem as an over counting problem in the following way.
The cross section of the virtual KK graviton exchanges $\sig$ is
proportional to the integration
\begin{eqnarray}
\sig \propto \int_{0}^{\alpha\Mf} \!\!\!\! dm_1 
  \!\! \int_{0}^{\alpha\Mf} \!\!\!\! dm_2 \  \rho(m_1) \rho(m_2)
        \M_\Gamma(m_1)\M_\Gamma^\dagger(m_2)\ , \label{eq.mass_int}
\end{eqnarray}
where the $\M_\Gamma(m)$ is an amplitude of the exchange process of 
the virtual KK graviton with mass $m$. 
Since the integration over mass corresponds to the phase space integration
in the extra dimensions, the double integration in
eq.(\ref{eq.mass_int}) includes a double counting of the phase space in
the extra dimensions. 
If there were no double counting 
(or no interference terms between KK gravitons), 
the problem would not appear in the virtual KK graviton exchange process. 
In fact, such double counting of the phase space never
appears in any process of the SM
because of the conservation law of the 4-momentum.

\subsection{Case of the process with more than two final states}

In the previous subsections, we discussed only 
the two-body scattering processes, but
in other process of the virtual KK graviton exchanges, 
for example
\begin{eqnarray*}
  &&pp \rightarrow {\rm jet} + G \\
    &&\hphantom{e^+e^- \rightarrow \gamma \ \ \ }
        \raise1.116ex\hbox{$\lfloor$}\2!\! 
    \xrightarrow[\hphantom{\textit{K-Kgrav}}]{} \  l\bar{l}  
\end{eqnarray*}
we can see the same problem.
In case that the number of final states is three,
it can be shown that
a fake enhancement of the on-shell contribution occurs under the
integration over mass, and the cross section becomes unphysically huge
compared to the cross section of that of the real KK graviton production
process, for example $ pp \rightarrow {\rm jet} + {\rm missing}$.

The cross section $\sig$ of the virtual KK graviton exchange processes
with the three final states, like  
$ pp \rightarrow ({\rm jet} + G) \rightarrow {\rm jet} + l\bar{l}$,
is expressed as
\begin{eqnarray}
  \sig &=& \d\sum_{ \ \left| \vec{n_1} \right|,\ 
   \left| \vec{n_2} \right| < \ \alpha\Mf R } \sig(\vec{n_1}, \vec{n_2})
      \nonumber \\
    &\thickapprox& \int_{0}^{ \alpha\Mf} \!\!\!\! dm_1 
  \!\! \int_{0}^{ \alpha\Mf} \!\!\!\! dm_2 
  \  \rho(m_1) \rho(m_2)\ \sig(m_1,m_2)\ . \label{eq:3body_full}
\end{eqnarray}
In eq.(\ref{eq:3body_full}), $\sig(m_1,m_2)$ stands for the value of the 
three-dimensional phase space integral of square of the  absolute value
of the amplitude of the process.
On the other hand, the cross section  $\sigma_{\rm real}$
 of the real KK graviton 
production processes which are included in such a virtual KK graviton
exchange process, like $ pp \rightarrow {\rm jet} + {\rm missing}$,   
is expressed as 
\begin{eqnarray}
 \sigma_{\rm real} &=&  \d\sum_{\ \n < \ \s R } \sigma_{\rm real}(\vec{n}) \ 
   \nonumber \\      
   &\thickapprox&  
     \int_{0}^{\s} \2! dm \ \rho(m)\  \sigma_{\rm real}(m)\ ,
\end{eqnarray} 
where $\sigma_{\rm real}(m)$ 
stands for the production cross section of the one 
KK graviton with mass $m$. 
We can separate $\sig$ into the two parts
\begin{eqnarray}
 \sig = \sig^{\rm m_1 \neq m_2} + \sig^{\rm m_1 = m_2}\ ,
              \label{eq:separate}
\end{eqnarray}
where $\sig^{\rm m_1 \neq m_2}$ and $\sig^{\rm m_1 = m_2}$ are 
given as follows:
\begin{eqnarray}
&&\4!\4!\4! \sig^{\rm m_1 \neq m_2} = \int_{0}^{\alpha\Mf} \!\!\!\! dm_1 
  \!\! \int_{0}^{m_1} \!\!\!\! dm_2 \  \rho(m_1) \rho(m_2)
  \ \sig(m_1,m_2)  \ ,\label{eq:cro_m1neqm2_def}
\\
&&\4!\4!\4! \sig^{\rm m_1 = m_2} = \int_{0}^{\alpha\Mf} \!\!\!\! dm_1 
  \!\! \int_{0}^{\alpha\Mf} \!\!\!\! dm_2 \  \rho(m_1) \rho(m_2)
  \ \sig(m_1,m_2)\delta(\left| \vec{n_1} \right| - \left| \vec{n_2}\right|)
                \nonumber \\
&& \ \ \ \ = \int_{0}^{\alpha\Mf} \!\!\!\! dm\ 
  \rho^2(m)\times\frac{1}{R}\ \sig(m,m) \ .\label{eq:int_cro_mm1}
\end{eqnarray}
In eq.(\ref{eq:separate}), $\sig^{\rm m_1 \neq m_2}$ 
represents the contribution from the interference
terms between KK gravitons whose mass are different and 
$\sig^{\rm m_1 = m_2}$ is composed of the two contributions: 
one from the real KK gravitations propagation
and from the interference terms between the KK gravitons 
whose mass are the same.
The $\sig^{\rm m_1 = m_2}$ gives dominant contribution for the $\sig$.
Changing the integration variable $m$ to the 
absolute value $\n$ of the KK index, $ \sig^{\rm m_1 = m_2}$ and
$\sigma_{\rm real}$ are expressed as
\begin{eqnarray}
 \sigma_{\rm real} &=& \int_{0}^{ \s R} \!\!\!\! d\n 
  \              \rho(\n) \ \sigma_{\rm real}(m)  \\
 \sig^{\rm m_1 = m_2} &=& \int_{0}^{ \alpha\Mf R} \!\!\!\! d\n 
  \  \rho^2(\n) \ \sig(m,m) \ ,  \label{eq:sig_n_int}
\end{eqnarray}
where
\begin{eqnarray}
 \rho(\n) = S_{\delta}\  m^{\delta -1}R^{\delta -1}
\end{eqnarray}
 is the multiplicity of the KK gravitons with mass $m = \n/R$,
and the integrand in eq.(\ref{eq:sig_n_int}) is proportional to the 
factor $\rho^2(\n)$, because all the KK gravitons with the same mass
$\n$ equally contribute,

Furthermore, $\sig^{\rm m_1 = m_2}$ is decomposed as 
the two contributions,
one from the KK graviton propagations (the diagonal part)
and one from the interference
terms between the KK gravitons whose mass are the same 
(the off-diagonal part): 
\begin{eqnarray}
\sig^{\rm m_1 = m_2} = \sig^{\rm diag} + \sig^{\rm off-diag}\ ,
\end{eqnarray}
where $\sig^{\rm diag}$ and $\sig^{\rm off-diag}$,
 the contributions from the KK gravitons
propagations and from the interference
terms, are given by
\begin{eqnarray}
 \sig^{\rm diag} &=& \d\sum_{ \ \left| \vec{n_1} \right|,\ 
   \left| \vec{n_2} \right| < \ \alpha\Mf R } \sig(\vec{n_1}, \vec{n_2})
             \ \delta^{ \vec{n_1},\vec{n_2} }  \nonumber \\
    &\thickapprox& \int_{0}^{ \alpha\Mf} \!\!\!\! dm\  \rho(m)
                          \sig(m,m) \ , \label{eq:cross_diag}\\
  \sig^{\rm off-diag}&\thickapprox& \int_{0}^{ \alpha\Mf} 
        \!\!\!\! dm\  \left(\rho^2(m) -\rho(m) \right) 
                          \sig(m,m) \ . \label{eq:cross_offdiag}
\end{eqnarray} 
We can regard $\sig(m,m)$ as the production cross section of one KK
graviton with mass $m$ because
\begin{eqnarray}
 \sig(m,m) = 
      \left\{ \begin{array}{ll}
   \sigma_{\rm real}(m)  \times 
    \left[ Br({\rm KK}\ {\rm decay}) 
      + \mathcal{O}\left( \frac{\Gamma(m)}{m} \right) \right] 
                            & ({\rm if} \ m \leq \s) \\
  \sigma_{\rm real}(m) \times \mathcal{O} \left( \frac{\Gamma(m)}{m} \right) 
                            & ({\rm if} \ m > \s)\ ,
\end{array} \right.  \label{eq:real_NWA}
\end{eqnarray} 
where $Br({\rm KK}\ {\rm decay})$ is a branching ratio of the decay
channel of the KK gravitons which is included in the full process.
The first equation in eq.(\ref{eq:real_NWA}) is exactly the expression 
of the narrow width approximation for the one KK graviton exchange
process.
Substituting eq.(\ref{eq:real_NWA}) into
eq.(\ref{eq:cross_diag}), we get
\begin{eqnarray}
 \sig^{\rm diag} \simeq \int_{0}^{ \s} \!\!\!\! dm\  \rho(m)
             \sigma_{\rm real}(m) \times Br({\rm KK}\ {\rm decay}) \ .
\end{eqnarray}
If this were the whole contribution, then the narrow width approximation
would be satisfied.
However, it can be shown that $\sig$ is dominated by the contribution 
from the interference terms: 
\begin{eqnarray}
 \sig \simeq \sig^{\rm off-diag} \simeq \int_{0}^{ \s R} \!\!\!\! d\n 
  \  \left( \rho^2(\n) - \rho(\n)\right) 
      \ \sigma_{\rm real}(m) \times Br({\rm KK}\ {\rm decay})\ ,
\end{eqnarray} 
and we get the ratio of the cross section of $\sig$ to $\sigma_{\rm real}$
\begin{eqnarray}
 \frac{\sig}{\sigma_{\rm real}} &\sim& \rho(\n = \s R)
                    \nonumber \\
       &=& \O \left(\frac{\mp^{2 -\frac{2}{\delta}}}
          {\Mf^{(2+\delta)\left(1
	     -\frac{1}{\delta}\right)}}\s^{\delta-1}\right)\ ,
\end{eqnarray}
where $\rho(\n = \s R)$ is the number of the KK gravitons with mass 
$m = \s$.

Thus we find that the narrow width approximation is broken.
This is because the interferences between the KK gravitons 
whose mass are same are mistaken for the propagations of  
the real KK gravitons, and the cross section of 
the virtual KK graviton exchange 
$\sig$ get to be huge compared to the KK graviton production 
cross section $\sigma_{\rm real}$.

As compared with the case of the two-body scattering, there is a difference in
the representation of the over counting problem.
In case the virtual KK graviton exchanges with the three final
states, the over counting appears not only in the virtual KK gravitons 
but also in the contribution from the on-shell KK gravitons production,
although in case of the two-body scattering, it appears in the most
heaviest KK gravitons production and on-shell contribution of the 
KK gravitons is canceled out.
This is because, in the case with the three final states,
the pole of the propagators of the KK gravitons are taken care of with 
the three-dimensional phase space integral (not the mass integration).
In the case with more than three final state, first we can take care of
the on-shell
pole of the KK gravitons with the three-dimensional phase space
integral, and calculate the on-shell contribution and the off-shell
contribution of the KK gravitons correctly.
Therefore, the on-shell contribution of the KK gravitons are not
canceled, and the over counting also appears in the contribution 
from the on-shell KK graviton productions.

In case that the number of final states is more than three,
we find
a similar fake enhancement of the on-shell contribution
and the cross section again becomes unphysically huge
compared to the cross section of that of the real KK graviton production
process.

\section{Summary and conclusion}

In this paper, we have pointed out that the
the cross sections of the virtual KK graviton exchanges
in the ADD model have pathological behaviors.
We have shown that the narrow width approximation is not valid 
for the cross section of the two-body scattering process via the
KK gravitons. The cross section of the two-body scattering process
 get to be unphysically small compared to that 
of the real KK graviton production
process which is included in such a process, and strongly depends on the
cutoff parameter $\alpha$.

The cause of this problem lies in the over counting of the phase space in
the extra dimensions due to the violation of the momentum 
conservation in the extra dimensions.
If there were no double counting, the problem would not appear 
in the virtual KK graviton exchange process.

To search for the ADD model (or the large extra dimensions)
with collider experiments,
we need more care in evaluation of the virtual KK graviton exchange process.

\section*{Acknowledgments}
I received generous support from O. Yasuda and N. Kitazawa.
I would like to thank T. Yamashita, M. B. Gavela, S. Matsumoto and
S. C. Park for insightful comments and suggestions.

\end{document}